\newif\iffigs\figsfalse

\documentstyle[12pt]{article}

\iffigs
  \input epsf
\else
\fi

\batchmode
  \newfont{\footscrfont}{rsfs10}
  \newfont{\footbbbfont}{msbm10}
\errorstopmode

\newif\ifscrf\scrftrue
\ifx\footscrfont\nullfont
  \scrffalse
\fi

\newif\ifamsf\amsftrue
\ifx\footbbbfont\nullfont
  \amsffalse
\fi

\def\ppnumber{IASSNS-HEP-94/9}
\def\ppdate{February 1994}
\def\pplogo{\vbox{\kern-\headheight\kern -1pt
\halign{##&##\hfil\cr&{
\ppnumber}\cr\rule{0pt}{2.5ex}&\ppdate\cr}
}}

\makeatletter
\date{}
\def\dedicatory#1{\def\@date{\normalsize\it#1}}
\def\subjclass#1{\def\@thefnmark{}\@footnotetext{1991
    {\it Mathematics Subject Classification.} #1}}
\def\keywords#1{\def\@thefnmark{}\@footnotetext{
    {\it Key words and phrases.} #1}}

\def\ps@firstpage{\ps@empty \def\@oddhead{\hss\pplogo}%
  \let\@evenhead\@oddhead 
}
\def\maketitle{\par
 \begingroup
 \def\thefootnote{\fnsymbol{footnote}}
 \def\@makefnmark{\hbox
 to 0pt{$^{\@thefnmark}$\hss}}
 \if@twocolumn
 \twocolumn[\@maketitle]
 \else \newpage
 \global\@topnum\z@ \@maketitle \fi\thispagestyle{firstpage}\@thanks
 \endgroup
 \setcounter{footnote}{0}
 \let\maketitle\relax
 \let\@maketitle\relax
 \gdef\@thanks{}\gdef\@author{}\gdef\@title{}\let\thanks\relax}

\def\abstract{\if@twocolumn
\section*{Abstract}
\else \small
\begin{center}
{\bf ABSTRACT}
\end{center}
\quotation
\fi}

\newif\iffn\fnfalse

\@ifundefined{reset@font}{\let\reset@font\empty}{} 
\long\def\@footnotetext#1{\insert\footins{\reset@font\footnotesize
    \interlinepenalty\interfootnotelinepenalty
    \splittopskip\footnotesep
    \splitmaxdepth \dp\strutbox \floatingpenalty \@MM
    \hsize\columnwidth \@parboxrestore
   \edef\@currentlabel{\csname p@footnote\endcsname\@thefnmark}\@makefntext
    {\rule{\z@}{\footnotesep}\ignorespaces
      \fntrue#1\fnfalse\strut}}}
\makeatother

\ifamsf
  \newfont{\bigbbbfont}{msbm10 scaled\magstep2}
  \newfont{\bbbfont}{msbm10 scaled\magstep1}  
  \newfont{\smallbbbfont}{msbm8}
  \newfont{\tinybbbfont}{msbm6}
  \newfont{\smallfootbbbfont}{msbm7}
  \newfont{\tinyfootbbbfont}{msbm5}
\fi

\ifscrf
  \newfont{\scrfont}{rsfs10 scaled\magstep1}  
  \newfont{\smallscrfont}{rsfs7}
  \newfont{\tinyscrfont}{rsfs7}
  \newfont{\smallfootscrfont}{rsfs7}
  \newfont{\tinyfootscrfont}{rsfs7}
\fi

\ifamsf
  \newcommand{\Bbb}[1]{\iffn
      \mathchoice{\mbox{\footbbbfont #1}}{\mbox{\footbbbfont #1}}
      {\mbox{\smallfootbbbfont #1}}{\mbox{\tinyfootbbbfont #1}}\else
      \mathchoice{\mbox{\bbbfont #1}}{\mbox{\bbbfont #1}}
      {\mbox{\smallbbbfont #1}}{\mbox{\tinybbbfont #1}}\fi}
\else
  \def\bigbbbfont{\bf}
  \def\Bbb{\bf}
\fi

\ifscrf
  \newcommand{\Scr}[1]{\iffn
    \mathchoice{\mbox{\footscrfont #1}}{\mbox{\footscrfont #1}}
    {\mbox{\smallfootscrfont #1}}{\mbox{\tinyfootscrfont #1}}\else
    \mathchoice{\mbox{\scrfont #1}}{\mbox{\scrfont #1}}
    {\mbox{\smallscrfont #1}}{\mbox{\tinyscrfont #1}}\fi}
\else
  \def\Scr{\cal}
\fi

\def\operatorname#1{\mathop{\rm #1}\nolimits}
\def\C{{\Bbb C}}

\def\P{{\Bbb P}}
\def\Q{{\Bbb Q}}
\def\R{{\Bbb R}}
\def\Z{{\Bbb Z}}

\def\Vol{\operatorname{Vol}}

\def\opeq#1{\advance\lineskip#1 \advance\baselineskip#1
	\advance\lineskiplimit#1}
\def\eqalign#1{\null\,\vcenter{\opeq{2.5\jot}\mathsurround=0pt
	\everycr={}\tabskip=0pt
	\halign{\strut\hfil$\displaystyle{##}$&$\displaystyle{{}##}$\hfil
	\crcr#1\crcr}}\,\null}

\def\sm{$\sigma$-model}

\def\CY{Calabi-Yau}

\def\cM{{\Scr M}}

\def\cMc{{\hfuzz=100cm\hbox to 0pt{$\;\overline{\phantom{X}}$}\cM}}

\def\ff#1#2{{\textstyle\frac{#1}{#2}}}

\begin{document}
\setcounter{page}0
\title{\LARGE Resolution of Orbifold Singularities\\ in String Theory\\[10mm]
\insert\footins{\hbox to\hsize{\footnotesize
To appear in ``Essays on Mirror Manifolds 2''.\hfil}}}
\author{\vbox{
\begin{tabular}{c}
\normalsize Paul S. Aspinwall\\[5mm]
\normalsize School of Natural Sciences\\
\normalsize Institute for Advanced Study\\
\normalsize Princeton, NJ  08540
\end{tabular}}
}

{\hfuzz=10cm\maketitle}

\def\Large{\large}
\def\LARGE{\large\bf}

\vskip 3cm		

\begin{abstract}

In this paper the relationship between the classical description of
the resolution of quotient singularities and the string picture is
discussed in the context of $N$=(2,2) superconformal field theories.
A method for the analysis of quotients locally of the form
$\C^d/G$ where $G$ is abelian is presented.
Methods derived from mirror symmetry are used to study the moduli
space of the blowing-up process. The case $\C^2/\Z_n$ is
analyzed explicitly.

\end{abstract}

\vfil\break

\section{Introduction}

One of the earliest manifolds considered for a superstring target space
was a blown-up orbifold \cite{CHSW:}. In this case the orbifold (the
``Z-orbifold'') was constructed by modding out a torus of 6 real
dimensions by a group isomorphic to $\Z_3$. This group action has 27
fixed points leading to 27 isolated quotient singularities in
the resulting space. Algebraic geometry then tells us that these
singularities can be smoothed away (in particular, ``blown-up'')
to give us a \CY\ manifold. In this
context the orbifold itself is little more than a step in the
construction of a smooth \CY\ manifold.

Shortly afterwards it was realized \cite{DHVW:} that string theory is
well-behaved (at least in some sense) on the orbifold itself before
the smoothing process. It was also noticed that calculations of such
quantities as the field-theory version of the Euler characteristic
when performed on the orbifold agreed with the same geometric
calculation on the target space {\em after\/} blowing up. It is as if
the string theory ``knows'' about the blow-up even before it has been
performed. Thus the Euler characteristic formula of \cite{DHVW:} can
be taken as one of the earliest indications of the power of
``stringy'' geometry. By ``stringy'' or ``quantum'' geometry we mean
geometry as indicated via field theory analysis of a string theory
rather than direct analysis of the geometry of the string's target
space.

In many respects toroidal orbifolds such as the Z-orbifold provide
attractive models for string target spaces. The torus itself is a
little too trivial whereas a general \CY\ manifold can render many
calculations very difficult. The toroidal orbifold may be thought of
as a good compromise since the field theory view of the quotient
singularities is rather straight-forward.
Having said this, proponents of \CY\ manifolds would say that the
blown-up version of the orbifold is more general --- the orbifold
itself is just a limiting case of the blown-up orbifold. Thus people
who study orbifolds are artificially singling out a special class of
string target spaces. We will argue below that this point of view
might be a little unfair.

Before discussing orbifolds any further we should first state clearly
what class of spaces we are studying. Orbifolds were introduced
many years ago \cite{Sat:V}\footnote{The original name was
``V-manifold''. The term ``orbifold'' appears to be due to
W.P.~Thurston.}. As originally defined an orbifold may be taken to be
a space (an algebraic variety)
whose only singularities are locally of the form of quotient
singularities. In the context of this paper we will only consider
``\CY\ orbifolds'', that is an orbifold which is a complex algebraic
variety of complex dimension $d$ with holonomy group $\subseteq
SU(d)$. The class of \CY\ orbifolds clearly covers the case of a
complex $d$-torus
divided by a discrete group $G\subset SU(d)$ and indeed any \CY\
$d$-fold divided by $G$.

As an example of an orbifold {\em not\/} of the form $M/G$ for $M$ a smooth
\CY\ manifold consider again the case of the Z-orbifold. Clearly the
Z-orbifold is of the form $M/G$ but if we blow-up just one of the 27
quotient singularities we destroy this property. The resultant space
still has 26 quotient singularities locally of the form $\C^3/\Z_3$
however and is thus an orbifold. String theory is still well-defined
on this space indicating that it is the local rather global form of the
singularities that is important. We therefore believe that the
mathematicians definition of an orbifold should also be the appropriate
definition in string theory.

In order to focus on purely local aspects of quotient singularities we
will restrict ourselves to isolated singularities in this paper. This
is done mainly for reasons of clarity and it should be
straight-forward to generalize most of what is said in this paper to
non-isolated quotient singularities. We also assume that any smooth
manifold appearing in this paper should be K\"ahler.

Let us now clarify the term ``blowing-up''. A simple example of a
blow-up, which will be the main focus of this paper, acts to alter
slightly a singular space so that it becomes smooth. The singularity
is thus resolved. A blow-up is a particular kind of resolution of
singularity and so we require more explanation of the term. It is
actually easier to describe the reverse process of
``blowing-down''. That is we take a smooth manifold and change it so
that it becomes singular.

Given a \CY\ manifold $X$ with a metric
there are two types of continuous changes we
consider. Firstly, deformation of complex structure of $X$
(i.e., what geometers would normally call simply a deformation) and
secondly deformation of the K\"ahler form on $X$. It turns out that
the blowing-down process corresponds to a deformation of the K\"ahler
form. The moduli space of K\"ahler forms on $X$ has a cone structure
as is easy to see --- if $g_{i\bar\jmath}$ is a valid Hermitian metric on
$X$ then so is $\lambda g_{i\bar\jmath}$ for $\lambda$ a positive real
number. As we approach the wall of the K\"ahler cone of $X$ from the
interior, some algebraically embedded subspace of $X$ or $X$ itself
will shrink down acquiring zero size on the wall. If it is a
subspace that shrinks down rather than the whole of $X$, we say that
this subspace has been {\em blown-down\/} when we reach the wall.
Thus the orbifolds in classical geometry which may be blown-up
into smooth \CY\ manifolds lie at the edge of
the K\"ahler cone as depicted in figure \ref{fig:oc}.

\iffigs
\begin{figure}
  \centerline{\epsfxsize=7cm\epsfbox{blow-oc.ps}}
  \caption{The K\"ahler cone of a resolved orbifold.}
  \label{fig:oc}
\end{figure}
\fi

At first sight the analysis of the string theory on an orbifold
appears to have a similar structure to that suggested by the
description of classical geometry above. One can build the $N$=(2,2)
superconformal
field theory corresponding to the orbifold and find the dimension of
the moduli space in which it lives by counting the truly marginal
operators preserving the $N$=(2,2) world-sheet supersymmetry
in the theory. When one does this one finds the correct
number for the dimension of the moduli space of the {\em maximally
resolved\/} orbifold. By maximally resolved we mean that the most
blow-ups have been done to resolve the singularities consistent with
the $K=0$ (vanishing first Chern class) condition. Thus it
appears that the some of the marginal operators in the orbifold conformal
field theory correspond to blow-up modes. Indeed, as we will show
later in this paper they are in natural one-to-one correspondence.

This is not the full story however. When one analyzes the structure of
the moduli space around an orbifold point, the conformal field theory
picture and the classical geometric picture do not appear to agree
\cite{me:orb1}. In particular it is not clear how a cone-like
structure appears in the moduli space spanned by the marginal
operators when one looks at a neighbourhood of the orbifold point.
This problem was effectively resolved in \cite{AGM:I,AGM:II,W:phase}
where it was shown how cone structures did appear naturally in the
moduli space of $N$=(2,2) superconformal
field theories as one moves sufficiently far
away from the orbifold point. This allows one to recover the classical
picture of a blow-up in the conformal field theory language once one
has moved sufficiently far in moduli space. The problem is that the
cone structure found in \cite{AGM:I,AGM:II,W:phase} is in complete
disagreement with the picture of the orbifold point in figure
\ref{fig:oc}!

The situation discovered in \cite{AGM:I,AGM:II,W:phase} may be stated
roughly as follows. Given a certain class of theories (corresponding
to \CY\ manifolds which may be realized as hypersurfaces in toric
varieties) the moduli space of suitably defined
K\"ahler forms (to be precise the
``algebraic measures'' of \cite{AGM:sd}) is isomorphic to
$\R^{h^{1,1}}$. One can now subdivide this space into a fan-like
structure consisting of cones with their apex at the origin of
$\R^{h^{1,1}}$. This decomposition divides the moduli space into
so-called ``phases''. One of these cones will be congruent to the
K\"ahler cone of the \CY\ manifold. The other cones appear to be best
thought of as associated to other geometries. One of the other
geometries is an orbifold. In such a picture, the orbifold point
itself lies in the deep interior of the orbifold cone.

\iffigs
\begin{figure}
  \centerline{\epsfxsize=10cm\epsfbox{blow-os.ps}}
  \caption{The stringy view of figure \protect\ref{fig:oc}.}
  \label{fig:os}
\end{figure}
\fi

In figure \ref{fig:os} we show how this phase picture modifies the
classical picture of figure \ref{fig:oc}. Only part of the fan is
shown for clarity.
In order to reach the
orbifold point one must leave the \CY\ K\"ahler cone.

The above state of affairs was clarified in \cite{AGM:sd}
by thinking about correlation functions arising from
string theory on a \CY\ manifold, $X$. The string
theory is given in terms of a non-linear \sm\ with target space $X$.
Some correlation functions in this theory are prone to instanton
corrections which correspond to rational curves. The precise form of
the instanton correction is as a power series in $q_j$ where
$q_j=\exp(-A_j)$ and $A_j$ is the area of a curve representing
the $j$th generator
of $H_2(X,\Z)$. If all the rational curves are sufficiently large then
this power series will converge (assuming a non-zero radius of
convergence). In this case there is no problem analyzing the \sm.
By varying the K\"ahler form on $X$, we can shrink some of the
rational curves on $X$. As we do this we may eventually move outside
the region of convergence of instanton power series. Now the analysis
of the \sm\ is unclear. It might happen however that the field theory
can be interpreted more simply in terms of a different target space.
This is exactly what happens in the case of an orbifold. When we are
in the orbifold phase, correlation functions may be determined in
terms of the orbifold theory perturbed in some way (actually by twist
fields, see for example \cite{Cve:orb}).
The orbifold cone marks roughly, in some way, the region of
convergence of this new perturbation theory.

Even though the phase picture naturally divides up the moduli space
into regions with different geometrical interpretations one should not
take the word ``phase'' too seriously --- a generic path from one
phase region into another will not encounter a theory that is singular
in any way. Thus one can choose
one geometrical interpretation and ``analytically continue'' it into
other regions. In terms of the correlation functions above this
analytic continuation is quite literal. Thus, for example, we might
describe the whole moduli space in terms of the \CY\ region. This is
what we will do later on this paper to determine the precise conformal
field theory picture of blowing-up.
This will allow us to make contact with figure \ref{fig:oc} by
redefining the K\"ahler form.
It should be emphasized however
that this is an artificially biased approach in favour of smooth
target spaces. One could equally choose the orbifold region to
analytically continue over the whole space. It is in this sense that
the \CY\ proponent has no claims being ``more general'' than the
orbifold proponent. It should also be noted that there are examples
where at least one of the proponents would fail, i.e., there are \CY\ manifolds
which cannot be blown down to orbifolds and orbifolds which cannot be
blown up into smooth \CY\ manifolds. It is therefore important to always
consider both methods of approach.

In section \ref{s:cl} we will review the classical theory of blow-ups
of singularities formed by quotients by abelian groups in a language
that will be useful later in the paper. In section \ref{s:tw} we will
look at the same problem from the conformal field theory point of
view and derive a direct map correspondence between the two
approaches.
In section \ref{s:ms} we will analytically continue the \CY\ region
into the orbifold region to measure the size of the blown-up parts of
the orbifold.
Finally in section \ref{s:ex} we will study some details of the structure of
the moduli space of blowing-up codimension 2 singularities.


\section{Classical Geometry}  		\label{s:cl}

Let us first outline the concept of a (K\"ahler) blow-up.
In much of what follows we will
consider the case where $X$ is not compact. In particular $X$ will
correspond to $\C^d/G$ for some group $G$. Since we are only
interested in the local behaviour of string theory around a quotient
singularity this will suffice.
Given a K\"ahler
manifold $X$, one can vary the sizes of algebraically embedded
submanifolds of $X$ by varying the K\"ahler form on $X$. In particular
one might have a divisor (i.e., an algebraically embedded codimension
one\footnote{Dimensions of the target space
will always be counting {\em complex\/}
degrees of freedom in this paper.} subspace) which can be shrunk down to an
arbitrarily small size while keeping other divisors (not in the same
homology class) at finite size. Such a divisor is {\em exceptional}.
If one shrinks an exceptional divisor down to zero size then the limit
of $X$ in this process may
be singular. We say then that $X$ has been blown-down along this
exceptional divisor. As we shall see, if one blows down along the
right combination of divisors then the resultant
singularity is a quotient singularity. Thus we have described the
reverse of the process of blowing-up a quotient singularity to obtain
a smooth manifold $X$. The singular point set in an orbifold when
blown-up appears to be replaced
by an exceptional divisor (which may be the sum of many irreducible
divisors) in $X$.

We will now review how any singularity locally in the form of $\C^d/G$
where $G$ is a discrete {\em abelian\/} group can be blown-up.
We will
focus on the situation of an isolated singularity (i.e., the origin
is the only fixed point of any non-trivial element of $G$) but the
following method also can be used in the non-isolated case.
We also want to consider the case in which we do not affect the canonical
class so that the resultant space, if smooth, will be a \CY\
manifold.
A simple holonomy argument shows that in this case, $G\subset SU(d)$.
To do the blowing-up
we will use the language of toric geometry. It is beyond the
scope of this article to explain toric geometry --- for that the
reader is referred to \cite{Fulton:} or \cite{AGM:II}. See also
\cite{MOP:} for an early account of this in the physics literature. To fully
understand the following arguments one should consult these references
but it is hoped that the following be self-contained if one requires
only a schematic understanding.

A toric variety $X_\Delta$ is a $d$-dimensional complex space
specified by a fan $\Delta$. In this paper, the fan $\Delta$ is a collection of
{\em simplex\/}-based cones (together with their faces) in $\R^d$. One also
puts a lattice structure in the same space, i.e., $\Z^d\subset\R^d$.
This lattice is denoted ${\bf N}$. $X_\Delta$ is compact if and only
if $\Delta$ spans the whole of $\R^d$. There is a
correspondence between $p$-codimensional algebraic subspaces of
$X_\Delta$ and $p$-(real)-dimensional subcones of $\Delta$. Thus,
divisors in $X_\Delta$ correspond to 1 dimensional cones, i.e., rays
in $\Delta$.

\iffigs
\begin{figure}
  \centerline{\epsfxsize=10cm\epsfbox{blow-tg.ps}}
  \caption{ A Toric Description of $X$.}
  \label{fig:tg}
\end{figure}
\fi

In figure \ref{fig:tg} we show an example of a fan
$\widetilde\Delta$ describing
a toric variety $X_{\widetilde\Delta}$.
(We use $\widetilde\Delta$ rather than $\Delta$ for this particular example to
fit in with notation used below.)
In this case $\widetilde\Delta$ lives in $\R^3$ and
consists of three 3-dimensional cones. The configuration of these
cones is shown by slicing $\widetilde\Delta$ by a hyperplane $\Pi$ as shown in
the figure. Since $\widetilde\Delta$ does not fill $\R^3$,
$X_{\widetilde\Delta}$ is not
compact. $X_{\widetilde\Delta}$ contains a divisor, $E$, which corresponds to
the ray indicated in the figure. In this case the other 3 rays
correspond to divisors not homologically distinct from $E$.

We now discuss the toric picture of blowing up. In the context of this
paper, a blow-up consists of adding a divisor with one or more
irreducible component into $X_\Delta$.
Thus we must add rays to the fan $\Delta$. This process is easily
visualized in terms of the intersection of $\Delta$ with $\Pi$. To add
an irreducible divisor to $X$,
draw a point in the interior of $\Delta\cap\Pi$ and
then draw lines from it to form a triangulation of the $(d-1)$-dimensional
simplex (or simplices) in which the point lies.
Thus the simplex containing the new point is subdivided into a set of
simplices each having this new point as a vertex. This
simple case is known as a
{\em star subdivision\/} of the fan. A sequence of many star-subdivisions
gives a blow-up whose exceptional divisor consists of many irreducible
divisors.
We can illustrate an
example of the process of a single star-subdivision
by the toric variety pictured in figure
\ref{fig:tg}. In this case we consider
$X_{\widetilde\Delta}$ to be the blown-up
toric variety from a variety $X_\Delta$. The exceptional divisor
is $E$. The process is shown in figure \ref{fig:b1} where we draw only
the intersection of the fan with $\Pi$ rather than the whole fan
itself.

\iffigs
\begin{figure}
  \centerline{\epsfxsize=10cm\epsfbox{blow-b1.ps}}
  \caption{Blowing up $X_\Delta$ with $E$ as an exceptional divisor.}
  \label{fig:b1}
\end{figure}
\fi

Now let us consider the questions of when $X_\Delta$ in singular and
when $X_\Delta$ has trivial canonical class. For both of these
questions we need to look at the lattice structure ${\bf N}$ and the
way in which it relates to $\Delta$. Consider following each ray in
$\Delta$ out from the origin of the fan, which is itself a lattice
point, until we reach another lattice
point. Mark this point in $\R^d$. The condition for trivial canonical class
is simple to state --- all these points must lie in a
hyperplane\footnote{This
hyperplane must also be distance ``one'' in suitable units from the origin
to ensure that $K=0$ rather than only $nK=0$ for some $n>1$. See, for
example, the discussion of $\Q$-Cartier divisors in \cite{Fulton:}.This
situation will not arise for the class of examples we consider here.}.
We will imagine $\Pi$ to be this plane from now on. Now consider each
simplex in $\Delta\cap\Pi$. Each $(d-1)$-dimensional simplex forms
a $d$-dimensional simplex by including the origin of the fan as
another vertex.
A volume (or length or area etc.) of this $d$-simplex may be
calculated where
the volume is defined\footnote{Note that the conventions in this paper
differ by a factor of $d!$ from those of \cite{AGM:II}.}
in terms of units defined by the
lattice ${\bf N}$.
$X_\Delta$ is smooth if and only if each such simplex
has volume $1/d!$. Thus we see from figure \ref{fig:b1} that if
$X_{\widetilde\Delta}$ is smooth then $X_\Delta$ is singular since the
tetrahedron subtended by the
triangle in $\Delta\cap\Pi$ has volume $\ff12\neq\ff1{3!}$.

Now all we need to complete our discussion of resolving quotient
singularities is the form in which $\Delta$ represents a
quotient singularity. It turns out that a singularity of the type
$\C^d/G$ for abelian $G$ is represented by a fan where
$\Delta\cap\Pi$ consists of just one $(d-1)$-dimensional simplex.
Suppose we consider the space $\C^d$ with coordinates $(z_1,z_2,
\ldots,z_d)$ and $X_\Delta\cong\C^d/G$ with coordinates
$(x_1,x_2,\ldots,x_d)$ away from the origin\footnote{To be more
precise, $(x_1,x_2,\ldots,x_d)$ are the coordinates on the {\em
algebraic torus}.}. Let $\vec\alpha_1,\vec\alpha_2,\ldots,
\vec\alpha_d$ be the
vertices of the simplex $\Delta\cap\Pi$ which have coordinates
\begin{equation}
\begin{array}{rcl}
  \vec\alpha_1&=&(a_{11},a_{12},\ldots,a_{1d})\\
  \vec\alpha_2&=&(a_{21},a_{22},\ldots,a_{2d})\\
  \multicolumn{3}{c}{\vdots}\\
  \vec\alpha_d&=&(a_{d1},a_{d2},\ldots,a_{dd}).
\end{array}
\end{equation}
The coordinates above are defined with respect to the lattice ${\bf
N}$, i.e., the points with integer coordinates are the points in ${\bf
N}$. This implies that the $a_{ij}$'s are integers given our
definition of $\Pi$ above.

We now state that the transformation between $X_\Delta$ and $\C^d$ is
given by
\begin{equation}
\begin{array}{rcl}
  z_1^{a_{11}}z_2^{a_{21}}\ldots z_d^{a_{d1}}&=&x_1\\
  z_1^{a_{12}}z_2^{a_{22}}\ldots z_d^{a_{d2}}&=&x_2\\
  \multicolumn{3}{c}{\vdots}\\
  z_1^{a_{1d}}z_2^{a_{2d}}\ldots z_d^{a_{dd}}&=&x_d.
\end{array}	\label{eq:r1}
\end{equation}
The above relationships are sufficient to determine $G$ for a given
set of $a_{ij}$'s. Suppose an element $g\in G$ acts on $\C^d$ as
\begin{equation}
  g\;:\;(z_1,z_2,\ldots,z_d)\mapsto(e^{2\pi ig_1}z_1,e^{2\pi ig_2}z_2,
	\ldots,e^{2\pi ig_d}z_d), 		\label{eq:gact}
\end{equation}
where $g_i$ are rational numbers, $0\leq g_i<1$. The relations
(\ref{eq:r1}) imply that
\begin{equation}
  \sum_i g_i a_{ij}\in\Z\qquad\forall j.	\label{eq:gint}
\end{equation}

We can now blow-up $X_\Delta$ by subdividing the simplex with corners
$\vec\alpha_i$ with points that lie in the intersection of this simplex
with ${\bf N}$. Such a blow-up will not affect the canonical class.
Such a point $\vec\beta_k$ is given by
\begin{equation}
  \eqalign{\vec\beta_k &= \sum_i h_i^{(k)}\vec\alpha_i,\cr
	\sum_i h_i^{(k)} &= 1,\cr
	0\leq h_i^{(k)} &<1,\cr} \label{eq:hkbb}
\end{equation}
and that $\vec\beta_k$ has integer coordinates. From (\ref{eq:gint}) we
see that each point $\vec\beta_k$ is given by $h^{(k)}=g\in G$ such that
\begin{equation}
  \sum_i g_i=1.  \label{eq:tmm1}
\end{equation}

As a simple example,
suppose $d=3$ and $\Delta$ is given by
\begin{equation}
\begin{array}{l}
  \vec\alpha_1=(3,-1,-1)\\
  \vec\alpha_2=(0,1,0)\\
  \vec\alpha_3=(0,0,1).
\end{array}
\end{equation}
This implies that $G$ is isomorphic to $\Z_3$ and is generated by
\begin{equation}
  \zeta\;:\;(z_1,z_2,z_3)\mapsto(\omega z_1,\omega z_2,\omega z_3),
\qquad \omega=e^{2\pi i/3},
\end{equation}
i.e., $\zeta_i=\ff13$ for $i=1,2,3$.
In fact figure \ref{fig:b1} shows the blow-up of this singularity. The
point added, $\vec\beta_1$ has coordinates $(1,0,0)$ and is given from
(\ref{eq:hkbb}) by $h^{(1)}=\zeta$.

As a more complicated example consider the case where $d=3$ and
$G\cong\Z_{11}$ and is generated by
\begin{equation}
  (\zeta_1,\zeta_2,\zeta_3) = (\ff1{11},\ff3{11},\ff7{11}).
\end{equation}
In this case
\begin{equation}
\begin{array}{l}
  \vec\alpha_1=(11,-3,-7)\\
  \vec\alpha_2=(0,1,0)\\
  \vec\alpha_3=(0,0,1).
\end{array}
\end{equation}
The points, $\vec\beta_k$,
added to $\Delta\cap\Pi$ to form the blow-up together with
the corresponding element of $G$ are
\begin{equation} \def\arraystretch{1.1}
\begin{array}{lcl}
  \vec\beta_1=(1,0,0) & \mbox{given by} & \zeta \\
  \vec\beta_2=(2,0,-1) & \mbox{\tt "} & \zeta^2 \\
  \vec\beta_3=(4,-1,-2) & \mbox{\tt "} & \zeta^4 \\
  \vec\beta_4=(5,-1,-3) & \mbox{\tt "} & \zeta^5 \\
  \vec\beta_5=(8,-2,-5) & \mbox{\tt "} & \zeta^8. \\
\end{array}	\label{eq:Z11w}
\end{equation}

\iffigs
\begin{figure}
  \centerline{\epsfxsize=10cm\epsfbox{blow-b2.ps}}
  \caption{A blow-up of a $\C^3/\Z_{11}$ singularity.}
  \label{fig:b2}
\end{figure}
\fi

A blow-up of this $\Z_{11}$ quotient singularity is shown in figure
\ref{fig:b2}. Note that there is more than one way to subdivide this
triangle to add in the points $\vec\beta_k$. This corresponds to the fact
that there is more than one way of blowing up this singularity. No
matter how we do the subdivision we will always divide the big
triangle $\Delta\cap\Pi$ into 11 triangles subtending volumes of
$\ff16$. This means that all the blow-ups completely resolve
the $\Z_{11}$ singularity to a
smooth space.

Note that in the case of the $\Z_{11}$, the complete resolution
required an exceptional divisor with 5 irreducible components which
correspond to $\vec\beta_1,\ldots,\vec\beta_5$. Each of these components is
independent with regards to homology. The precise geometry of
these 5 divisors depends on the triangulation of $\Delta\cap\Pi$ and
figure \ref{fig:b2} shows just one possibility.

It is a matter of combinatorics to show that this process of
blowing up will always fully resolve a singularity of the form
$\C^3/G$ where $G\subset SU(3)$ is abelian.
It is interesting to note however that
this is not the case for more than 3 dimensions. Consider the case of
$\C^4/\Z^2$ given by the generator
\begin{equation}
  \zeta\;:\;(z_1,z_2,z_3,z_4)\mapsto(-z_1,-z_2,-z_3,-z_4).
\end{equation}
In this case
\begin{equation}
\begin{array}{l}
  \vec\alpha_1=(2,-1,-1,-1)\\
  \vec\alpha_2=(0,1,0,0)\\
  \vec\alpha_3=(0,0,1,0)\\
  \vec\alpha_4=(0,0,0,1).
\end{array}
\end{equation}
The tetrahedron $\Delta\cap\Pi$ subtends a 4-simplex with the origin
with volume $\ff1{12}$ implying, as
expected, that $X_\Delta$ is singular, but this time there are {\em
no\/} points in ${\bf N}$ inside this tetrahedron on which one might
subdivide. Thus there is no toric resolution of this singularity
preserving vanishing canonical class. In fact D.~Morrison
\cite{M:priv} has shown that the results of \cite{Fine:} may be used
to show that there is no resolution
of this singularity of {\em any\/} kind which preserves $K=0$.

Let us now discuss the way in which the K\"ahler form controls the
blow-up. The K\"ahler form, $J$, may be expanded as
\begin{equation}
  J=\sum_{\xi=1}^{h^{1,1}} J_\xi\, e_\xi,	\label{eq:Jex}
\end{equation}
where $J_\xi\in\R$ and
the $e_\xi$'s form basis of $H^2(X,\Z)$. Imposing the condition that
all curves, surfaces, etc., have positive volume will put restrictions
on the allowed range of values of $J_\xi$ to give the K\"ahler cone.
In the simple case of a curve having homology class $\sum
c_\xi e_\xi$, the area will be $\sum J_\xi c_\xi$.

We may also calculate the size of any divisor, $D_l$, as follows.
Suppose that $D_l$ is
a generator of $H_{2(d-1)}(X,\Z)$ and is dual to a cycle with homology
class $e_l$. We then have (with
suitable normalizations of volume)
\begin{equation}
\eqalign{
  \Vol(D_l) &= \int_{D_l} J^{\wedge(d-1)}\cr
	&=\sum_{n_1,n_2,\ldots} J_{n_1}J_{n_2}\ldots
	\int_{D_l} e_{n_1}\wedge e_{n_2}\wedge\ldots\cr
	&=\sum_{n_1,n_2\ldots} (D_l\cap D_{n_1}\cap D_{n_2}\cap\ldots)\,
	J_{n_1}J_{n_2}\ldots,\cr
}
\end{equation}
where $(D_1\cap D_2\cap\ldots\cap D_d)$ represents the intersection
form on $X$. Thus we see that the volume of a divisor is given by an
expression more complicated than that for curves and it
depends on the intersection numbers.

Suppose we take a smooth manifold, $X$, and blow-down an exceptional divisor
to form an isolated quotient singularity. Let $D_{r_1},D_{r_2},\ldots$
represent the irreducible components of the exceptional divisor which
we assume are some of the generators of $H_{2(d-1)}(X,\Z)$ and are
thus dual to $e_{r_1},e_{r_2},\ldots\in H^2(X,\Z)$. The part of
$H_{2(d-1)}(X,\Z)$ not spanned by $D_{r_1},D_{r_2},\ldots$ may be
thought of as the homology which descends from the covering space of
our quotient singularity and this has no intersection with
$D_{r_1},D_{r_2},\ldots$. Thus we see that for $J_{r_1}=J_{r_2}=\ldots
=0$ we have $\Vol(D_{r_1})=\Vol(D_{r_2})=\ldots=0$. Thus the orbifold
is described by points corresponding to $J_{r_1}=J_{r_2}=\ldots=0$ on
the edge of the K\"ahler cone. It is important to realize however that
the parameters $J_{r_1},J_{r_2},\ldots$ do {\em not\/} independently control
the sizes of $D_{r_1},D_{r_2},\ldots$.

As an example consider a $\C^3/\Z_5$ singularity. There are two
components of the exceptional divisor, $A$ and $B$, such that
\begin{equation}
\begin{array}{ll}
  (A\cap A\cap A) = 9\qquad & (B\cap B\cap B)=8\\
  (A\cap A\cap B) = -3 & (A\cap B\cap B) = 1.
\end{array}
\end{equation}
Let us put $J=J_0 + ae_a + be_b$, where $J_0$ is the part of the
K\"ahler form given by cohomology elements away from the quotient
singularity.
We then have
\begin{equation}
  \eqalign{\Vol(A) &= (3a-b)^2\cr
  \Vol(B) &= (a-2b)(-3a-4b).\cr}   \label{eq:sD}
\end{equation}
We also have an algebraic curve homologous to $A\cap B$ which has area
$b-3a$ and another curve within $B$ with area $a-2b$.
Thus we have constraints on the K\"ahler cone $a-2b>0$, and
$b-3a>0$. These are sufficient to give positive volumes to $A$ and
$B$.

Finally in this section let us introduce the notion of a ``dual
curve''.  We have seen that given a component $D_l$ of the exceptional
divisor, the associated K\"ahler form gives the volume of the divisor
itself in terms of an expression like (\ref{eq:sD}).  If there were a
curve in $X$ whose homology class was dual to that of $D_l$, the
description of its area in terms of the K\"ahler form expansion would
be more direct.  Thus when thinking of performing a blow-up by
switching on components of the K\"ahler form it is probably best to
think in terms of the dual curves growing rather than the exceptional
divisor itself. This will be implicit in much of what follows. Note
that $D_l$ might be any element of $H_{2(d-1)}(X,\Z)$ and so in general
the dual must be thought as a rational combination of algebraic
curves. In the $\C^3/\Z_5$ example above, the K\"ahler cone imposes
the condition $a,b<0$. Thus there are no actual curves dual to $A$ or
$B$ since they would have negative area.


\section{Conformal Field Theory}   \label{s:tw}

In this section we will review and study the quotient singularities from the
point of view of the conformal field theory of string propagating on
the singular target space. The reader is referred to \cite{DFMS:} for
example for a more complete discussion. We should also point out that
we will not consider the case of nontrivial ``discrete torsion'' in
the sense of \cite{Vafa:tor} in this paper.

The field theory on $M/G$ is studied in terms of the theory on the
smooth covering space $M$. In addition to the $G$-invariant states
of $M$, the $M/G$ theory also has ``twisted'' states. These
may be thought of as open strings in $M$ whose ends are identified
by the element $g\in G$. Such a state is said to be in the
$g$-twisted ``sector''. As we shall see, it is the twisted states
which correspond to blow-up modes.

Let us analyze the chiral ring of the conformal field theory which is
expected \cite{LVW:} to
correspond to the cohomology ring of the target space. It is not too
hard to see that the $G$-invariant elements of the cohomology of $M$
will be precisely the untwisted modes on $M/G$ and so the twist fields
are somehow ``extra'' cohomology on $M/G$. Since the counting of
chiral fields should not change under a deformation of a theory, the
count for a blown-up orbifold should be the same as the orbifold
itself. Thus, the chiral fields which are twist fields must count
elements of cohomology which are generated during the blowing-up
process.

In conformal field theory language, the deformations are done by truly
marginal operators which, in the context of $N$=(2,2) theories, takes
a particularly simple form. That is, an action $S_0$ may be deformed
into $S$ as
\begin{equation}
  S = S_0 + \sum_k a_k\int\Phi_k\,d^2\theta^+d^2z
	+ \sum_r \tilde a_r\int\widetilde\Phi_r\,d\bar\theta^+d\theta^-d^2z
	+ \mbox{h.c.}		\label{eq:defm}
\end{equation}
The $\Phi_k$ and $\widetilde\Phi_r$ are (anti)chiral superfields with
$U(1)$ charges (left,right) equal to $(1,1)$ and $(-1,1)$
respectively. As is well-known, the \sm\ interpretation of these
superfields are as elements\footnote{The choice of signs of these
charges is a matter of convention.} of $H^{1,1}$ and $H^1(T)$ and thus as
deformations of K\"ahler form and complex structure respectively.
In order to focus on blow-ups, which are to be regarded as
deformations of the K\"ahler form, we will concentrate on the fields
$\Phi_k$.

Let us consider the spectrum of chiral twist-fields in an orbifold
$\C^d/G$ with only the origin as an isolated fixed point. Consider an
element $g\in G$ which acts on $\C^d$ with coordinates
$(z_1,z_2,\ldots,z_d)$ as in (\ref{eq:gact}) subject again to the
condition that $0\leq g_i<1$. One can then show (see, for example,
\cite{Vafa:Lorb} or \cite{Zas:}) that the $g$-twisted state associated
with the origin has charge $(Q,\bar Q)$ where
\begin{equation}
  Q=\bar Q = \sum_i g_i.
\end{equation}
Thus, in order to obtain twisted marginal operators associated to
blow-up modes we require precisely equation (\ref{eq:tmm1}) from the
previous section. That is, the classification of twisted marginal operators
associated to blow-ups has exactly matched the determination of
irreducible components of the exceptional divisor.

We have thus arrived at what might be called the
``(twist-field)-divisor'' map. Since the classifications are
isomorphic one can naturally identify a natural one-to-one correspondence.
E.g., for the $\Z_{11}$ example one may
consider figure \ref{fig:b2} and equation (\ref{eq:Z11w}) where the
last column in (\ref{eq:Z11w}) now represents the group element by which
the twist-field is twisted.

If we consider the $\C^4/\Z_2$ singularity in the last section which
had no $K=0$ resolution then we see that in the language of conformal
field theory, there are no twisted marginal operators associated with
the fixed point. Thus geometry and conformal field theory again agree.

Actually what we have done thus far is little more than paraphrase the
work of \cite{AGM:II,AGM:mdmm} where the ``{\em
monomial\/}-divisor mirror map'' was introduced. The r\^ole of
monomials in a Landau-Ginzburg theory has been replaced by twist
fields in the mirror of that theory. In fact, it can be seen from the
example studied in \cite{AGM:II} that the equivalence of monomials in
a Landau-Ginzburg theory and twist-fields in the mirror arises
naturally and can be traced back to the work of \cite{GP:orb}. All we
have done in this paper is to free the formalism of
\cite{AGM:II,AGM:mdmm} from references to the global space as a whole so
we need concentrate only on the quotient singularities locally.

We can now continue our study of the conformal field theory of
blow-ups by continuing to copy results from the monomial-divisor
mirror map to the (twist-field)-divisor map. In particular we can
study quantitatively the relationship between the coefficients $a_k$
in (\ref{eq:defm}) and exactly the size and shape of the exceptional
divisor.

The question of determining the size and shape of the exceptional
divisor immediately forces one to face questions to do with quantum
verses classical geometry. One knows that if the coefficients $a_k$
are small then the exceptional divisor will be small and we are thus
dealing with lengths possibly near the Planck scale.
This is the point at which we must address the issues raised in the
introduction concerning the phase picture of the moduli space. We are
going to try to interpret the orbifold phase in terms of sizes of
exceptional divisors and thus are biasing ourselves in favour of the
smooth \CY\ phase.
In order to
do this we must begin in the \CY\ phase, i.e.,
when the singularity is blown-up so that all radii are large and we
can make contact with classical notions of length.

Within the \CY\ phase there is a natural definition of K\"ahler form.
This is given by the ``\sm\ measure'' defined in
\cite{AGM:sd}. It cannot be emphasized too strongly that this is {\em not\/}
the same K\"ahler form that was used in the introduction to draw
figure \ref{fig:os}. The K\"ahler form used in the introduction was
abstractly defined and treats each phase equally. See \cite{AGM:sd}
for a more complete discussion of these two K\"ahler forms. The
K\"ahler form derived from the \sm\ arises as follows.
Suppose, $d=3$ and, for simplicity, that $H^{2,0}(X)\cong0$. Let us
consider the 3-point function between three
chiral fields $\Phi_l$, $\Phi_m$ and $\Phi_n$. Assuming we have a
purely geometrical interpretation, $X$, of the theory, we can associate
these fields to elements of $H^{1,1}(X)$ and thus divisors $D_l$,
$D_m$ and $D_n$. It can then be shown that \cite{DSWW:,DG:exact,AM:rat}
\begin{equation}
  \langle\Phi_l\Phi_m\Phi_n\rangle= (D_l\cap D_m\cap D_n)
    +\sum_\Gamma\frac{{\bf q}^\Gamma}{1-{\bf q}^\Gamma}(D_l\cap\Gamma)
    (D_m\cap\Gamma)(D_n\cap\Gamma),			\label{eq:3p}
\end{equation}
where $\Gamma$ is a holomorphically embedded $\P^1$
 in $X$ and ${\bf q}^\Gamma$ is a monomial
in the variables $q_\xi$.
We define the
parameters $q_\xi$ by
\begin{equation}
  q_\xi = \exp\{2\pi i(B_\xi+iJ_\xi)\},
\end{equation}
where the K\"ahler form, $J$, on $X$ has been expanded as in
(\ref{eq:Jex}).
The antisymmetric tensor,
``$B$-field'', on $X$ is also an element of $H^2(X,\R)$ on $X$ and is
similarly expanded to give $B_\xi$. We have
implicitly used the same normalization $4\pi^2\alpha^\prime=1$ as in
\cite{AGM:sd}.

In the neighbourhood of a large radius limit, the expansion
(\ref{eq:3p}) is sufficiently single-valued to allow a determination
of $J$ from (\ref{eq:Jex}). This then allows us to work out the size
of any divisor or, to be more precise,
the area of the dual curve as in section \ref{s:cl}.
This is the sense in which the conformal field theory data, i.e., the
three point functions $\langle\Phi_l\Phi_m\Phi_n\rangle$ can be used
to ``measure'' sizes within $X$.


\section{Exploring the Moduli Space}		\label{s:ms}

We will now discuss the relationship between the coupling constants
$a_k$ in (\ref{eq:defm}) and the parameters $J_l$. This will tell us
the precise way in which the twisted marginal operators perform the
blow-up of the orbifold.
In general there is a very complicated relationship between the
$a_k$'s and $J_l$'s but in this paper we will study the situation when
only one component is added to the exceptional divisor at a time. This
is the picture illustrated in figure \ref{fig:sq}. In terms of toric
geometry this resolution consists of a sequence of star-subdivisions
of the fan corresponding to the orbifold. One should note that not all
resolutions can be generated by a sequence of star-subdivisions and
that we have thus lost some generality in the following discussion. The
reader might wish to verify that the resolution in figure \ref{fig:b2}
can be obtained by a sequence of star-subdivisions, e.g.,
$\vec\beta_2,\vec\beta_3,
\vec\beta_1,\vec\beta_4,\vec\beta_5$.

\iffigs
\begin{figure}
  \centerline{\epsfxsize=14cm\epsfbox{blow-sq.ps}}
  \caption{The resolving a singularity in steps.}
  \label{fig:sq}
\end{figure}
\fi

Consider the situation in which
\begin{equation}
  S = S_1+\sum_u a_u \int\Phi_u\,d^2\theta^+d^2z,
		\label{eq:LGm}
\end{equation}
and suppose each field $\Phi_u$ corresponds to a ray in a fan
$\Delta$. Let the label $l$ denote which point we are going to add
to give our star subdivision and let
the field $\Phi_l$ be the corresponding twisted marginal
operator. In this case there will be a plane $\Pi$ as discussed in the
previous section intersecting this ray and its neighbours (but not
necessarily all the rays in the fan). Associate the position vectors
$\vec\alpha_u$ as the points of intersection between $\Pi$ and each
ray. There will then be some minimal relationship
\begin{equation}
  \sum_{u\in U_l} N_u\vec\alpha_u - N_l\vec\alpha_l = \vec0,
		\label{eq:star}
\end{equation}
for the set of rays $U_l$ which give the vertices of the minimum
dimension simplex in $\Pi$ which is being star-subdivided.
The integers $N_u,N_l$ are positive and have highest
common factor 1. We then define
\begin{equation}
  z_l = (-a_l)^{-N_l}\prod_{u\in U_l} a_u^{N_u}.
\end{equation}
The hypothesis of the monomial-divisor mirror map \cite{AGM:mdmm} then
tells us \cite{AGM:sd} that
\begin{equation}
  (B+iJ)_l = \ff1{2\pi i}\log z_l + O(z_l),  \label{eq:mdmm}
\end{equation}
for $|z_l|\ll1$.

The above expansion is expected to converge for
$|z_l|\ll1$, that is, $J_l\gg0$. This is the sense in which we have
had to go to the large-radius limit to make contact with classical
geometry. Note that, as one would expect, this limit is achieved by
adding in a large amount of twist field to the action, i.e., having a
large coefficient $a_l$.

The situation explored in \cite{AGM:mdmm} was for the mirror of a
Landau-Ginzburg theory. In this situation the form (\ref{eq:LGm}) made
sense --- each $\Phi_u$ corresponds to a monomial in the
superpotential of the mirror. In a general orbifold theory, it is not
clear how to write an action in the form (\ref{eq:LGm}). Since we
expect the analysis of a blow-up to not depend on global properties
of a manifold however we claim that some version of the above analysis
can be performed for any abelian quotient singularity. That is, we may
obtain some ``normalized'' variable $z_l$ which controls the size of a
blow-up at the large radius limit in terms of the coefficient of the
marginal operator $a_l$. One should also note that it is by use of the
mirror map that we are effectively putting a connection on the bundle
of fields over the moduli space. When a marginal operator is used to
perturb a theory a finite distance in moduli space, one should
specify the way in which the marginal operator itself is changing
along this path. In the context of Landau-Ginzburg theories, one
can write down a superpotential in terms of parameters --- the
coefficients within the superpotential. By varying these parameters
one moves a finite distance in moduli space. If one identifies fields
with the monomials in the superpotential one is also implicitly
defining the way in which these fields transform along the path. Thus
in this paper we are asserting that the twist-fields transform the same way
in which the monomials do in the Landau-Ginzburg-type theory.

Once we have the relationship (\ref{eq:mdmm}) we may use the local
geometry of the moduli space to determine the form of the
$O(z_l)$ term. Again we will use mirror symmetry to motivate our
argument. It was conjectured in \cite{CDGP:} and demonstrated in
\cite{BCOV:big} that given a pair of mirror spaces, $X$ and $Y$, with
the K\"ahler form on $X$ being given by the above notation then
\begin{equation}
  (B+iJ)_l = \frac{\int_{\gamma_l}\Omega}{\int_{\gamma_0}\Omega},
		\label{eq:mmap}
\end{equation}
where $\Omega$ is a (3,0)-form on $Y$ and $\gamma_0,\gamma_l$ are
elements of $H_3(Y,\Z)$. In the case that $Y$ is represented by a
Landau-Ginzburg theory with superpotential (\ref{eq:LGm}) we can
consider a period $\int_{\gamma}\Omega$ to be a function of the variables
$a_u$. In this case, the periods satisfy a set of coupled linear Fuchsian
partial differential equations known as the {\em Picard-Fuchs\/}
equations. In \cite{Bat:var} it was shown that this set of
differential equations were of the hypergeometric type studied by
Gel'fand et al in \cite{GG:h,GZK:h}. However, in \cite{GZK:h} these
differential equations were written down purely in terms of the toric
data of $X$. This means that in determining the periods as a function
of $a_u$, no direct reference need be made to the geometry of $Y$ ---
only that of $X$. This is exactly what we want in the context of this
paper since we are not directly interested in the structure of $Y$.

The Picard-Fuchs equations will have many solutions but the
singularity structure of the equations dictates that the solutions are
classified by their behavior near $z_l=0$. In fact, the equation
(\ref{eq:mdmm}) is sufficient to determine precisely which two periods
are required for the ratio in (\ref{eq:mmap}).

We are now in a position to track the entire process of a blow-up of a
singularity. The general solution of the Picard-Fuchs equation is
rather awkward to handle and so we will study only the decoupled
situations when we need only consider ordinary differential equations
(although it should be noted that a full multiparameter system can been
studied \cite{lots:per,Drk:Z,CDFKM:I,HKTY:}). It was shown in
\cite{AGM:sd} that if all the components of $J$ were held at $\infty$
except for one, $J_l$, the the Picard-Fuchs equations decoupled leaving an
ordinary differential equation of which two of the solutions would
provide the ratio for (\ref{eq:mmap}). It is also the case that if a
point is not included in the triangulation of $\Delta\cap\Pi$ then
that point plays no r\^ole in the formulation of the Picard-Fuchs
equations. Thus, if we consider the process of a star-subdivision on a
point, we may consider this process decoupled from the rest of the
system both before the star-subdivision is included (since then the
point is not in the triangulation) and when the large $J$ limit
of this process has been taken.

In summary then we consider the following process. Consider an
orbifold which is at its large radius limit, i.e., all components of
the K\"ahler form not associated with blow-ups are infinite. Then
blow-up the singularities by a sequence of processes as follows:
take a single star-subdivision
and slowly take that component of the K\"ahler form (i.e., the area of
the dual curve) to infinity. In
this way only one component of the K\"ahler form will be finite at any
given time and we will only ever have to deal with ordinary differential
equations.

We now state the ordinary differential equations required (see
\cite{AGM:sd} for more details). Given a star-subdivision on
$\vec\alpha_l$ given by (\ref{eq:star}) then
\begin{equation}
\eqalign{
\Biggr\{\prod_{u\in U_l}N_u^{N_u}(z_l\frac{d}{dz_l}-\frac{N_u-1}{N_u})
  (z_l\frac{d}{dz_l}&-\frac{N_u-2}{N_u})\ldots
  (z_l\frac{d}{dz_l}-\frac1{N_u})z_l\frac{d}{dz_l}\cr
  -z_lN_l^{N_l}(z_l\frac{d}{dz_l}+\frac{N_l-1}{N_l})
  (z_l\frac{d}{dz_l}&+\frac{N_l-2}{N_l})\ldots
  (z_l\frac{d}{dz_l}+\frac1{N_l})z_l\frac{d}{dz_l}\Biggl\}f(z_l)=0,\cr}
		\label{eq:hyp1}
\end{equation}
where $f(z_l)$ is the required period. It is trivial to show that
$f(z_l)=1$ is a solution and we take this to be the period in the
denominator of (\ref{eq:mmap}).

We can write down another solution if we impose that one of the
$N_u$'s is equal to 1. Let us denote the set $U_l$ with this element
removed by $\bar U_l$. With a little effort, using the ideas in
\cite{GZK:h} and that $\sum_{u\in U_l}N_u=N_l$, one can show that
\begin{equation}
  \eqalign{
    h_l(z)&=\frac{N_l}{2\pi i}\int_{-i\infty}^{+i\infty}
	\frac{\Gamma(N_ls)\Gamma(-s)}{\displaystyle{\prod_{u\in\bar U_l}}
	\Gamma(N_us+1)}(-z)^s\,ds - \pi i\cr
    &= \log z + \sum_{n=1}^\infty \frac{N_l(N_ln+1)!}
	{n!\displaystyle{\prod_{u\in\bar U_l}}(N_un)!} z^n\cr}
		\label{eq:hz}
\end{equation}
is a solution.
The integration path is taken along the imaginary line deformed around
the origin to the left and the series is obtained by completing this
path to the right if we impose the condition $\arg(-z)<\pi$.
Hence from (\ref{eq:mdmm}) we deduce that the precise
K\"ahler form is given by
\begin{equation}
  (B+iJ)_l = \ff1{2\pi i}h_l(z_l).
\end{equation}

The region of convergence of the series in (\ref{eq:hz}) is given by
$|\hat z|<1$ or $\hat z=1$ where
\begin{equation}
  \hat z = \frac{N_l^{N_l}}{\displaystyle{\prod_{u\in U_l}
	N_u^{N_u}}}z.
\end{equation}

We can {\em analytically continue\/} $h_l(z)$ into the region $|\hat z|>1$ in
the usual way by closing the integration path in (\ref{eq:hz}) to the
left. We then obtain
\begin{equation}
  h_l(z) = -\pi i -\frac{e^{-\frac{\pi i}{N_l}}\Gamma(\frac1{N_l})}
	{\displaystyle{\prod_{u\in\bar U_l}}\Gamma(1-\frac{N_u}{N_s})}\,
	 \psi_l + O(\psi_l^2),		\label{eq:hzc}
\end{equation}
where $\psi_l^{-N_l}=z_l$ and $0<\arg(\psi_l)<\frac{2\pi}{N_l}$. Note
that if we normalize $a_u=1$ for $u\in U_l$, then we have
$\psi_l=a_l$. This means we can think of $\psi_l$ as the twist-field
coupling constant.

Without the blow-up, $a_l=0$ and thus $\psi_l=0$. This means that
before blowing up we have
\begin{equation}
  B_l=-\ff12,\qquad J_l=0.
\end{equation}

This value of $J_l$ agrees with classical geometry. That is, the
boundary of the K\"ahler cone at which $J_l=0$ is precisely where the
classical orbifold lives. This may be thought of in terms of the
commutativity of the following diagram:
\begin{equation}
\def\mapright#1{\smash{\mathop{\longrightarrow}\limits^
	{\hbox{\scriptsize #1}}}}
\def\mapdown#1{\Big\downarrow\rlap{$\vcenter{\hbox{\scriptsize #1}}$}}
\def\arraystretch{1.5}
\begin{array}{ccc}
	M&\mapright{\sm}&S_M\\
	\mapdown{/G}&&\mapdown{/G}\\
	M/G&\mapright{``\sm''}&S_M/G
\end{array}
		\label{eq:com1}
\end{equation}
In this diagram, $M$ is a smooth \CY\ manifold, $G$ is a symmetry by
which we can mod out and $S_M$ represents the conformal field theory
associated to $M$. The quotients taking us from the top row to the
bottom row are effected geometrically on the left and in terms of
field theory (i.e., twist-fields) on the right. The arrow on the top
line represents a \sm\ interpretation which is straight-forward
for large $M$ whereas the arrow in the bottom line represents the
``analytic continuation'' of a \sm\ to small radii (of exceptional
divisor).

It is important to notice that we have only established the
commutativity of (\ref{eq:com1}) when the orbifold is at large radius
limit away from the quotient singularities. That is, all the
components of the K\"ahler form associated to untwisted elements of
$H^2(X)$ are infinite. It is an interesting question to see if
commutativity still holds away from this limit. That is, given a
finite sized orbifold, do we get $J_l=0$ for the K\"ahler form
analytically continued from the large blow-up limit. This is addressed
in \cite{me:min-d}.

It is not yet clearly understood what the geometrical meaning of the
value $B_l=-\ff12$ at the orbifold point is.

Let us discuss more carefully the way in which we have performed
the analytic continuation in obtaining (\ref{eq:hzc}). The Fuchsian
differential equation (\ref{eq:hyp1}) has singularities at $\hat z_l=0,1,
\infty$ and thus we expect non-trivial monodromy of the solutions
around these points. We may think of $z_l$ (together with
$z_l=\infty$) as parametrizing a $\P^1$ in the compactified moduli
space of conformal field theories. We show this $\P^1$ in figure
\ref{fig:p1}.

\iffigs
\begin{figure}
  \centerline{\epsfxsize=16cm\epsfbox{blow-p1.ps}}
  \caption{The part of moduli space parametrized by $z_l$.}
  \label{fig:p1}
\end{figure}
\fi

If we wish to ascribe a single value of $(B+iJ)_l$ to each point on the
$\P^1$, we need to define branch cuts. Around $z_l=0$ we have infinite
monodromy familiar from the symmetry $B_l\cong B_l+1$ of the quantum
field theory. This tells us that there should be one branch line
coming out from this point. We effectively perform this cut by
imposing $-1<B_l<0$ which may be thought of as the $\arg(-z)<\pi$
condition implicit in (\ref{eq:hz}). The other end of this branch cut
is $\hat z_l=1$ --- the point at the edge of the convergence of the
series. The analytic continuation then implies a branch cut from $\hat
z_l=1\rightarrow z_l=\infty$. This cut was imposed by the condition
$0<\arg(\psi_l)<\frac{2\pi}{N_l}$.
These branch cuts are shown in figure
\ref{fig:p1}.

One would naturally expect the smallest values of $J_l$ to be close to
the orbifold point $\psi_l=0$. From (\ref{eq:hzc}) however we see that
for small values of $\psi_l$, our branch cut imposes the condition
$J_l\geq0$. This indicates strongly that $J_l\geq0$ over the entire
$\P^1$. It also indicates that all the $J_l$'s may be non-negative
over the whole moduli space.
This may be checked in simple examples but there is no general proof
of this fact yet. Thus it would appear that no measurements of
negative distance (area, etc.) are possible in this scenario.

The monodromy around $z_l=\infty$ is finite of order $N_l$.
We may therefore remove the branch cut locally by taking a finite cover
branched around this point in the moduli space.
Thus the
orbifold point may be thought of as a $\Z_{N_l}$-quotient singularity
itself\footnote{Our analysis does not determine the full codimension
of this singularity.
One should remember that codimensional one quotient
singularities may be removed by coordinate redefinition.} within the
moduli space. Note that the structure of the singularity in the moduli
space is generally different from the the singularity in the
target space although there are similarities. There will be a
$\Z_{N_l}$-quotient singularity in moduli space
for each irreducible component of the exceptional divisor
of the target space quotient singularity.

These orbifold points in the moduli space may also be seen to arise
in terms of ``quantum symmetries'' \cite{Vafa:qu}. It is known that
for string theory with a target space in the form of an orbifold $M/G$
where $G$ is a cyclic group, then $G$ is part of the symmetry group of
the theory and orbifolding by this group will retrieve the string with
target space $M$. The symmetry group $G$ acts nontrivially on the
twist fields and thus on the tangent bundle of the moduli space at the
orbifold point giving a quotient singularity in the moduli space. Note
however that this argument relies on global symmetries. If we return
to the example of the Z-orbifold, we have a $\Z_3$ quantum symmetry.
This is destroyed if any twist-field marginal operator is switched on
to resolve one of the 27 quotient singularities. Thus if we consider
switching on a second twist-field to blow-up another fixed point we
cannot use the quantum symmetry argument to find the structure of the
moduli space in the vicinity of this region. Our analysis of the
differential equations above however tells us that there is indeed a
$\Z_3$ quotient singularity in the moduli space for each of the 27 blow-ups.


\section{Quotient Singularities in Two Dimensions}
	\label{s:ex}

\begin{figure}
\setlength{\unitlength}{0.25mm}
$$\begin{picture}(480,280)(160,520)
\thinlines
\put(215,695){\makebox(0,0)[lb]{\raisebox{0pt}[0pt][0pt]{$\vec\beta_1$}}}
\put(255,695){\makebox(0,0)[lb]{\raisebox{0pt}[0pt][0pt]{$\vec\beta_2$}}}
\put(390,695){\makebox(0,0)[lb]{\raisebox{0pt}[0pt][0pt]{$\vec\beta_5$}}}
\put(300,695){\makebox(0,0)[lb]{\raisebox{0pt}[0pt][0pt]{$\vec\beta_3$}}}
\put(345,695){\makebox(0,0)[lb]{\raisebox{0pt}[0pt][0pt]{$\vec\beta_4$}}}
\put(240,680){\circle*{10}}
\put(280,680){\circle*{10}}
\put(320,680){\circle*{10}}
\put(360,680){\circle*{10}}
\put(400,680){\circle*{10}}
\put(200,520){\line( 1, 0){440}}
\put(160,680){\line( 1, 0){460}}
\put(200,520){\line( 1, 4){ 60}}
\put(200,520){\line( 1, 2){120}}
\put(200,520){\line( 3, 4){165}}
\put(200,520){\line( 1, 1){225}}
\put(200,520){\line( 5, 4){275}}
\thicklines
\put(200,800){\line( 0,-1){280}}
\put(200,520){\line( 3, 2){330}}
\put(630,670){\makebox(0,0)[lb]{\raisebox{0pt}[0pt][0pt]{\LARGE\char'05}}}
\end{picture}$$
  \caption{The fan for the resolution of $\C^2/\Z_6$.}
  \label{fig:c2}
\end{figure}
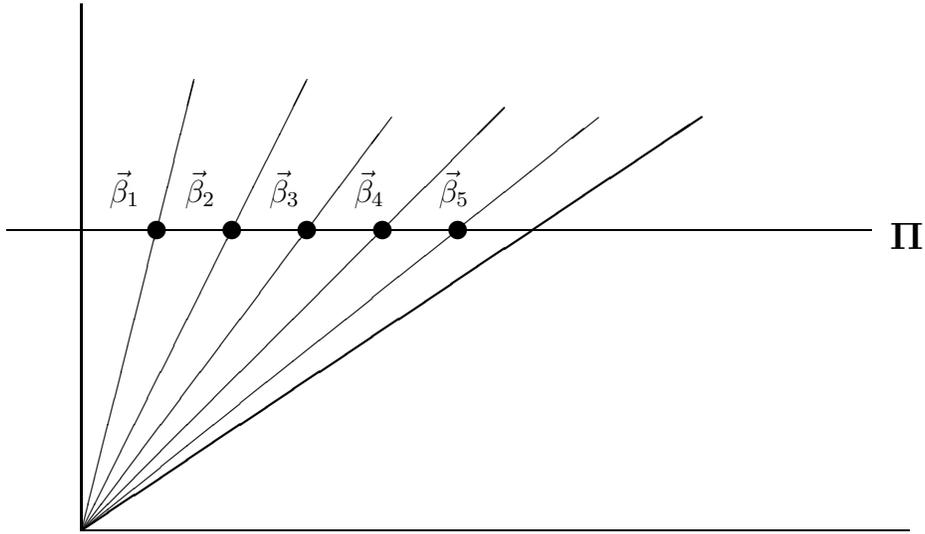

In this section we will illustrate the results of the last few
sections by considering the simplest set of quotient singularities
--- namely those of the form $\C^2/\Z_n$ given by the action
\begin{equation}
  \zeta\;:\;(z_1,z_2)\mapsto(e^{2\pi i/n}z_1,e^{-2\pi i/n}z_2).
\end{equation}
This corresponds to a matrix
\begin{equation}
  {\bf A}=\left(\begin{array}{cc}n&1\\0&1\end{array}\right).
\end{equation}
The lattice points in $\Delta\cap\Pi$ are thus given by
\begin{equation}
  \vec\beta_m = (m,1),\qquad m=1,\ldots n-1,
\end{equation}
corresponding to marginal operators in the $\zeta^m$-twisted sector.
The subdivision of the fan $\Delta$ to smooth this quotient is unique
and is shown (for $n=6$) in figure \ref{fig:c2}

The geometric interpretation of the blow-up in figure \ref{fig:c2} is
of a chain of $\P^1$'s each one touching it's neighbour at one point.
This corresponds to the well-known Hirzebruch-Jung string for the
resolution of this singularity (see for example \cite{BPV:}). Another
way of viewing this is as follows. Begin with a $\C^2/\Z_n$ quotient
and star subdivide on $\vec\beta_1$. One can show that the resulting
blow-up has an exceptional divisor of $\P^1$ (corresponding to
$\vec\beta_1$) and a singularity locally of the form $\C^2/\Z_{n-1}$ lying
on this $\P^1$. We can now blow-up this singularity with $\vec\beta_2$ to
give another $\P^1$, touching the first one, this time with a
$\C^2/\Z_{n-2}$ quotient singularity. This process is repeated giving
a chain of $n-1$ $\P^1$'s for which the space is smooth.

In order to find the twist-field version of the resolution of
$\C^2/\Z_n$ we thus need only study the star-subdivision on
$\vec\beta_1$ since the complete blow-up may be viewed in terms of a
sequence of such events for decreasing $n$. The relationship
(\ref{eq:star}) then becomes
\begin{equation}
  (n-1)\vec\alpha_1 + \vec\alpha_2 = n\vec\beta_1.
\end{equation}
Assuming our twist-field is normalized correctly, which would amount
to set $a_1=a_2=1$ in the Landau-Ginzburg theory of the mirror, we may
call the coupling constant in front of the marginal operator $\psi$, in
which case
\begin{equation}
  z = \psi^{-n},
\end{equation}
and
\begin{equation}
(B+iJ)_1 = -\frac{n}{4\pi^2}\int_{-i\infty}^{+i\infty}
	\frac{\Gamma(ns)\Gamma(-s)}{\Gamma(ns-s+1)}(-z)^s\,ds
	-\ff12,
\end{equation}
which may be expanded as
\begin{equation}
(B+iJ)_1 = \frac{1}{2\pi i}\left(\log z +\sum_{p=1}^\infty
	\frac{n(np+1)!}{p![(n-1)p]!}\,z^p\right),	\label{eq:qqz}
\end{equation}
for $|z|<(n-1)^{n-1}/n^n$ or as
\begin{equation}
(B+iJ)_1 = -\ff12 + \ff1{2\pi}e^{(\frac12+\frac1n)\pi
i}\psi+O(\psi^2),		\label{eq:qqpsi}
\end{equation}
for $|\psi|<n/(n-1)^{\frac{n-1}n}$ and $0<\arg(\psi)<2\pi/n$.

It was shown in \cite{AGM:sd} that for $n=2$
the K\"ahler form could be given in
closed form as
\begin{equation}
  (B+iJ)_1 = \frac{i}\pi\cosh^{-1}\frac1{2\sqrt{z}}=
	-\frac1\pi\cos^{-1}\frac\psi2.
\end{equation}

Finally, it is interesting consider the value of $J_1$ at the ``phase
transition'' point between the orbifold and the \CY\ manifold. That
is, the point at which string theory becomes singular and we are on
the boundary of the region of convergence of (\ref{eq:qqz}) and
(\ref{eq:qqpsi}). This occurs for $z=(n-1)^{n-1}/n^n$. In table
\ref{tab:t} we give $J_1$ for the first few values of $n$. The value
of $B_1$ at this point is 0 for any $n$. It is
interesting to note that for $n=2$, the \CY\ phase extends all the way
down to zero distance and so the orbifold theory at $\psi=0$ and the
phase transition point differ only in the value of $B_1$. This is no
longer the case for $n>2$ and the size of the exceptional divisor for
the singular theory gets progressively larger as $n$ increases.

\begin{table}\begin{center}
\begin{tabular}{|c|c|}
  \hline
  $n$ & $J_l$ at transition\\
  \hline
  2 & 0 \\
  3 & 0.11 \\
  4 & 0.18 \\
  5 & 0.22 \\
  \hline
\end{tabular}\end{center}
  \caption{Values of the K\"ahler form for $z=(n-1)^{n-1}/n^n$.}
  \label{tab:t}
\end{table}

\section*{Acknowledgements}

It is a pleasure to thank my collaborators Brian Greene and David
Morrison for useful conversations and the work on which much in this
paper is based. I would also like to thank S. Katz for some useful comments.
The author was supported by an NSF grant PHYS92-45317.

\end{document}